\documentclass[12pt]{iopart}

\usepackage{graphicx}
\usepackage{float,verbatim,amssymb,stmaryrd}
\usepackage{iopams}  
\usepackage{epsfig}

\begin{document}
	
	\title{Extremely intense laser-based electron acceleration in a plasma channel}
	
	\author[cor1]{Marija Vranic}
	\address{GoLP/Instituto de Plasmas e Fus\~ao Nuclear, Instituto Superior T\'ecnico, Universidade de Lisboa, 1049-001 Lisbon, Portugal}
	\ead{marija.vranic@ist.utl.pt}

	\author{Ricardo A. Fonseca$^{1,2}$}
	\address{$^1$GoLP/Instituto de Plasmas e Fus\~ao Nuclear, Instituto Superior T\'ecnico, Universidade de Lisboa, 1049-001 Lisbon, Portugal}
	\address{$^2$DCTI/ISCTE - Instituto Universit\'ario de Lisboa, 1649-026 Lisboa, Portugal}
	\ead{ricardo.fonseca@ist.utl.pt}
	
	\author{Luis O. Silva}
	\address{GoLP/Instituto de Plasmas e Fus\~ao Nuclear, Instituto Superior T\'ecnico, Universidade de Lisboa, 1049-001 Lisbon, Portugal}
	\ead{luis.silva@ist.utl.pt}

	\begin{abstract}
Laser pulses of extreme intensities ($I>10^{22}~ \mathrm{W/cm^2}$) are about to become available in the laboratory. 
The prepulse of such a laser can induce a plasma expansion that generates a low-density channel in near-critical gas jets. 
We present a study of channel formation and subsequent direct laser acceleration of electrons within the pre-formed channel. 
Radiation reaction affects the acceleration in several ways. 
It first interferes with the motion of the return current on the channel walls.  
In addition, it reduces the radial expelling efficiency of the transverse ponderomotive force, leading to the radiative trapping of particles near the channel axis. 
These particles then interact with the peak laser intensity and can attain multi-GeV energies.

	\end{abstract}
	
	%
	%
	%
	%
	%

\section{Introduction}

Plasmas can sustain accelerating fields on the order of $\mathrm{TeV/m}$ \cite{Malka_science_2002}, which makes them an exellent candidate for building small-scale accelerators \cite{Dawson}. 
Multiple laboraties around the world have obtained reproducible, quasi-monoenergetic electron beams in laser-produced wakefields. 
Electron energy as high  as 4 GeV has been reported from an all-optical experiment using this technology \cite{Leemans_4gev}.
Apart from the laser-wakefield acceleration, other ideas are being explored to make use of the collective plasma dynamics for particle acceleration. 
One of the alternative approaches is to accelerate electrons within a hollow plasma channel.  
Low-density plasma channels within a dense background can serve as a guide for laser light. 
This was recently exploited to guide the electrons between two wakefield acceleration stages \cite{Leemans_plasma_lens}.

Channels are particularly suited for experiments where laser transport over several milimeters of plasma is required. 
This concept has also been proposed using a laser-formed channel for inertial confinement fusion within a three-step igition scheme \cite{Tabak_POP}, where the first stage is the target compression and in the second stage a long ($\gtrsim10$ ps) laser forms a hollow channel in the coronal plasma. 
The pre-formed channel provides a path for an intense ignition laser that subsequently propagates  towards the target core. 
Due to the interaction at a lower density,  the laser penetrates through the coronal plasma with minimal losses of energy.
This enables a higher coupling efficiency of the laser energy to the ignition core.  
The potential global impact of this idea  has motivated further research on plasma channel formation and dynamics \cite{ pipe_channel, channel_formation1, Fuchs_channel, Malka_prl_1997_ch, Zulf_POP_2003, channel_electrons, Channel_formation_2, sarkisov, Sarri_popch,  Li_simulations, Satya_channel, Lemos_Grismayer_Dias}. 
Two-pulse configuration studies showed that not only the laser, but the accelerated electron beams can also be guided within a plasma channel \cite{double_pulse_guiding, double_pulse_guiding_2}.
Optimising the laser self-focusing  \cite{Self_focusing_early1, Self_focusing_early2} and self-guiding  is of particular importance for the laser channeling \cite{Mori_selffocus, naumova_poleff, Matsuoka}. 
Besides self-focusing, a range of other interesting collective effects were observed, both in experiments and numerical simulations.
Among them are the long-lived postsolitons that persist long after the laser-plasma interaction is completed \cite{Sarri_solitons, Louise_channel, Macchi_ionmotion, DKPOP}.

More recently, laser-generated channels gained attention as an enviroment by itself well-suited for particle acceleration. 
Various mechanisms have been reported as responsible for electron acceleration within the channels, where the betatron resonance is one of the most explored \cite{pukhov_acc_pop, mangles_intenseaccmech}. 
The channels are shown to sustain a large-scale azimuthal magnetic fields \cite{Ashley_Bfield} where the electrons perform betatron oscillations.
A considerable enhancement of the axial momentum and the total electron energy can be achieved via amplification of betatron oscillations \cite{POP_arefiev_2014, PRL_arefiev_2012, Huang_recent}. 
The existence of transverse fields also helps keeping the momentum of forward-moving electrons aligned with the channel axis. 
Self-generated electromagnetic fields can thus assist direct laser acceleration within the channel and allow energy gain beyond the vacuum acceleration limit \cite{Tsakiris_acc}.
Fluctuations of the longitudinal electric field affect the dephasing between the electrons and the laser, 
which in turn allows for generation of "superponderomotive" electrons within the channel \cite{Superponderomotive} (that can be similarly generated in solid-density plasmas  \cite{Sorokovikova_PRL}).
In addition, electron acceleration within the channel can be initially assisted by the presence of a surface wave on the channel walls. 
Here, the surface wave pre-accelerates the electrons, which are then further accelerated by direct laser acceleration or betatron resonance  \cite{Naseri_acc, Naseri_PRL_SW}. 
A comprehensive review of the factors that contribute to the electron acceleration within the channel can be found in Ref. \cite{JPP_arefiev_2015} and scaling laws in Ref. \cite{Khudik_scalings}. 
A distinguishing feature for prospective applications is that particles accelerated within plasma channels emit high frequency radiation that can be used as an X-ray source, 
which was demonstrated in recent experiments   \cite{Silvia_nat, Betatron_channel, Kneip_prl, Albert_channel_xrays}.
At higher laser intensities,  $\gamma$-rays can be emitted  \cite{Arefiev_PRL2016_gamma_rays}, and the radiation reaction becomes relevant for the electron dynamics. 

Next generation laser facilities will provide extreme intensities ($I>10^{24}~ \mathrm{W/cm^2}$) \cite{facilities_ELI, facilities_apollon, facilities_xcels}. 
These short pulses of intense light ($20 - 150$ fs)  will be preceeded by a long low intensity pedastal (or prepulse).
When interacting with a plasma slab, the prepulse can produce a hollow plasma channel, because its duration is longer than the characteristic timescale for the plasma ions (the pedastal duration can be on the order of nanoseconds).
A plasma channel can also be produced by a separate laser pulse, with a moderate intensity of $10^{18}~\mathrm{W/cm^2}$ and duration on the order of $10$ ps.
If the prepulse forms a channel, the main laser will be naturally aligned with the channel axis, and therefore its interaction with the plasma will happen within the pre-formed channel. 
Due to the high laser intensity, this interaction may be radiation reaction-dominated. 
It was reported from simulation studies that radiation reaction can affect the collective plasma dynamics in laser interaction with solid targets ($\gtrsim10^{22}$ cm$^{-3}$) \cite{Arefiev_PRL2016_gamma_rays, Pukhov_rrtrapp}. 
An investigation on how an extremely intense laser will interact with dense gas targets is still missing. 
This is timely because hydrogen gas jets are now available with densities up to $n  \simeq10^{21}$  cm$^{-3}$ \cite{dense_gas_jets}, and they will be employed in experiments with the next generation of lasers. 

Here we present a study of a mm-scale channel formation in underdense plasma and subsequent intense laser propagation through that channel. 
We demonstrate that stable, non-fully-cavitated channels can be created in plasmas of various densities. 
In particular, we studied the channel formation in the background plasma between $n_e=0.001~n_c$ and $n_e=0.1~n_c$, where $n_c$ is the critical density.
We consider a channel formation with a laser of $I=10^{18}~\mathrm{W/cm^2}$. 
Large-scale self-generated electric and magnetic fields are studied with 2D and 3D particle-in-cell simulations (full-scale 3D simulations were performed at low plasma density).  
We focus on the channel structure obtained from simulations at $n_e=0.1~n_c$, where the full cavitation is not achieved and 
the plasma density near the channel axis is $n_e=0.02~n_c$. 
We then consider different intense lasers ($\tau_0\sim ps$) propagating through this light pipe, and their interaction with the plasma. 
Radiation reaction is shown to play an important role already at $a_0=100$ (where the normalized vector potential $a_0$ is defined as $a_0=0.85\sqrt{I[10^{18}~\mathrm{W/cm^2}]}\lambda_0[\mathrm{\mu m}]$ for a linearly polarised laser). 
Qualitative differences are observed in the plasma dynamics with and without accounting for the electron energy loss due to high-frequency photon emission. 
There are differences in the channel wall dynamics, but the most striking difference is that the radiative trapping enables electrons to experience longer interactions with the laser, which in turn leads to a higher energy gain due to the direct laser acceleration. 
This also increases the number of accelerated electrons.
Without optimising the channel width and laser focusing parameters, we show that with a 10 PW, 150 fs laser one can obtain an electron beam with over 1.6 nC of super-ponderomotive electrons, with a 6 GeV energy cutoff in a 1.8 mm plasma channel.
For the same laser parameters, we also consider a case of propagation through a channel with a spatialy-varying width.
Increasing the laser power, or varying the channel properties to obtain laser guiding at a higher local laser intensity is expected to substantially increase the electron cutoff energy. 
For example, energy of 15 GeV can be obtained with $a_0=600$ for a propagation length of 0.5 mm. 
 
This paper is organised as follows. 
In the next section, we study the channel formation in near-critical plasmas using long ($\tau_0>$ 10 ps), weakly relativistic ($a_0\simeq1$) laser pulses. 
The space-charge field structure within the self-created chennels is discussed, as well as its strength as a function of the background plasma density.
Section 3 explores the propagation of intense laser pulses ($a_0\gtrsim100$, $\tau_0\sim1~$ps) through such preformed channels. 
Special attention is devoted to exploring the distinct plasma dynamics when radiation reaction becomes relevant. 
This is followed by an example relevant for the near-future 10 PW laser facilities ($a_0~100$, $\tau_0\sim150~$fs).
Finally, we present our counclusions.

 \section{Plasma channel generation; electromagnetic field structure within a pre-formed plasma channel}

 \begin{figure}
	\centering
	\includegraphics[width=1.0\textwidth]{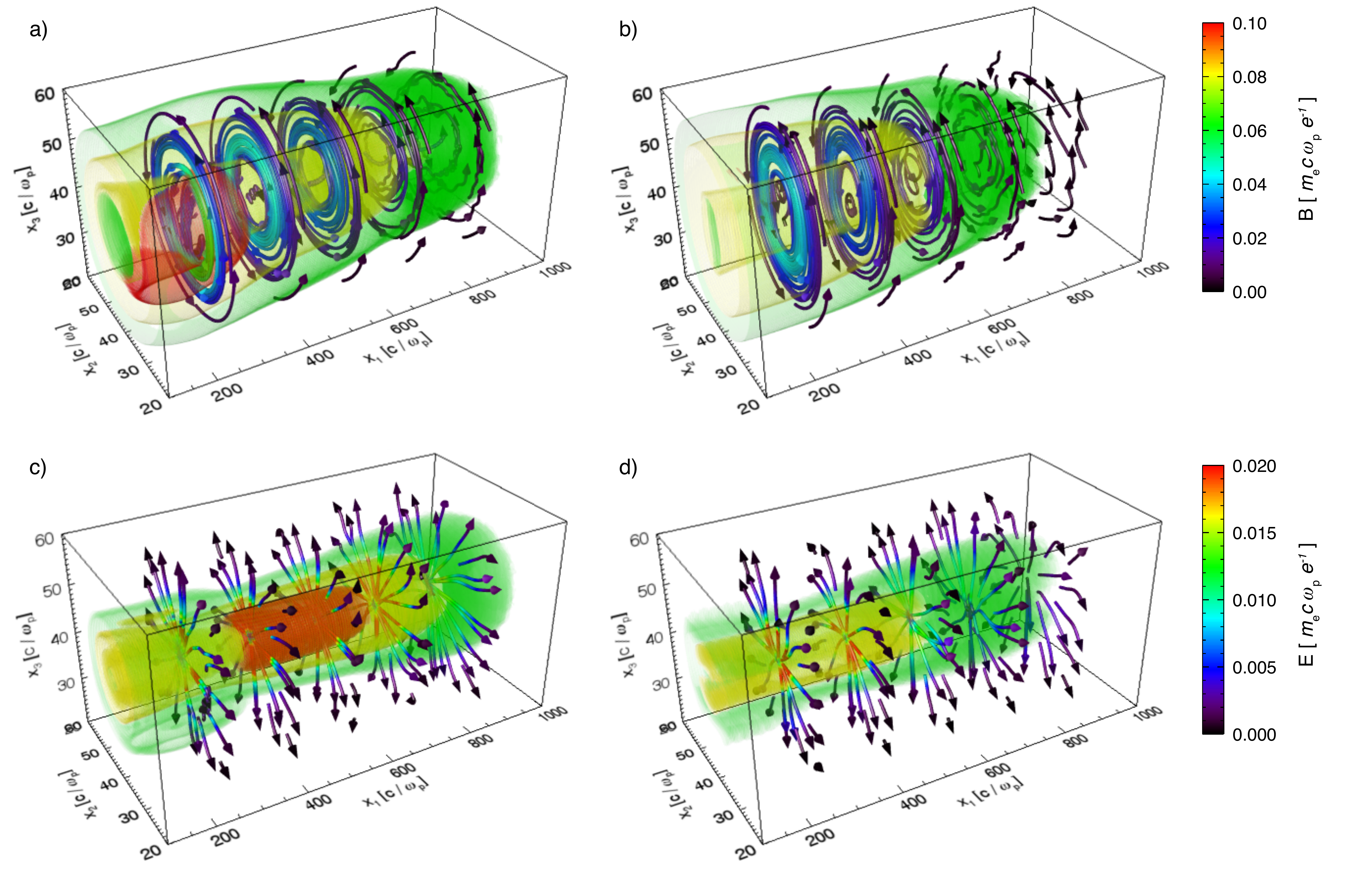}
	\caption{Electromagnetic field lines within a plasma channel formed for $n_p=0.001~n_c$ at $t=1632~ \omega_p^{-1}$ (9.07 ps). Magnetic field in a channel formed by a) circularly polarised laser, b) linearly polarised laser. Electric field in a channel formed by c) circularly polarised laser, d) linearly polarised laser.} 
	\label{3d_field_lines}
\end{figure}

This section deals with  $\gtrsim$10 ps long lasers interacting with underdense plasma slabs.
Such a long laser pulse creates a mm-scale plasma channel that can serve as a guiding structure for another laser beam. 
The laser field initially introduces a trasverse temperature gradient which causes the expansion of the electrons.  
Multi-ps timescale allows the plasma ions to follow the electrons, pulled by the electric field induced by the charge separation. 
If the laser power is high enough, self-focusing can increase the laser intensity, which further reinforces the channel formation. 
The critical power for self-focusing is given by  $ P_c=17 \left( n_c/n_e\right)  ~\mathrm{GW}$  \cite{Self_focusing_early1, Self_focusing_early2}, where $n_e$ is the electron plasma density and $n_c=m_e \omega_0^2/(4\pi e^2)$ is the critical density associated with the propagation of a light wave with frequency $\omega_0$. 
When the plasma density is near-critical, channel-splitting might occur, and a fraction of the laser energy may be lost trasversely. 
Self-focusing helps these daughter-channels to re-combine in a single main channel. 
If the side-channels persist, some of the energy can be trapped later to form solitons \cite{Sarri_solitons, Louise_channel, Macchi_ionmotion, DKPOP}. 
More details on channel formation process, laser self-focusing, self-guiding and associated instabilities can be found in the literature \cite{DKPOP, Friou_Gremillet, naumova_poleff, Naumova_popch, sarkisov, Mori_selffocus, clayton_selffocus, Mori_morecomplicated, Matsuoka, filament_ins}. 

 \begin{figure*}
	\centering
	\includegraphics[width=0.8\textwidth]{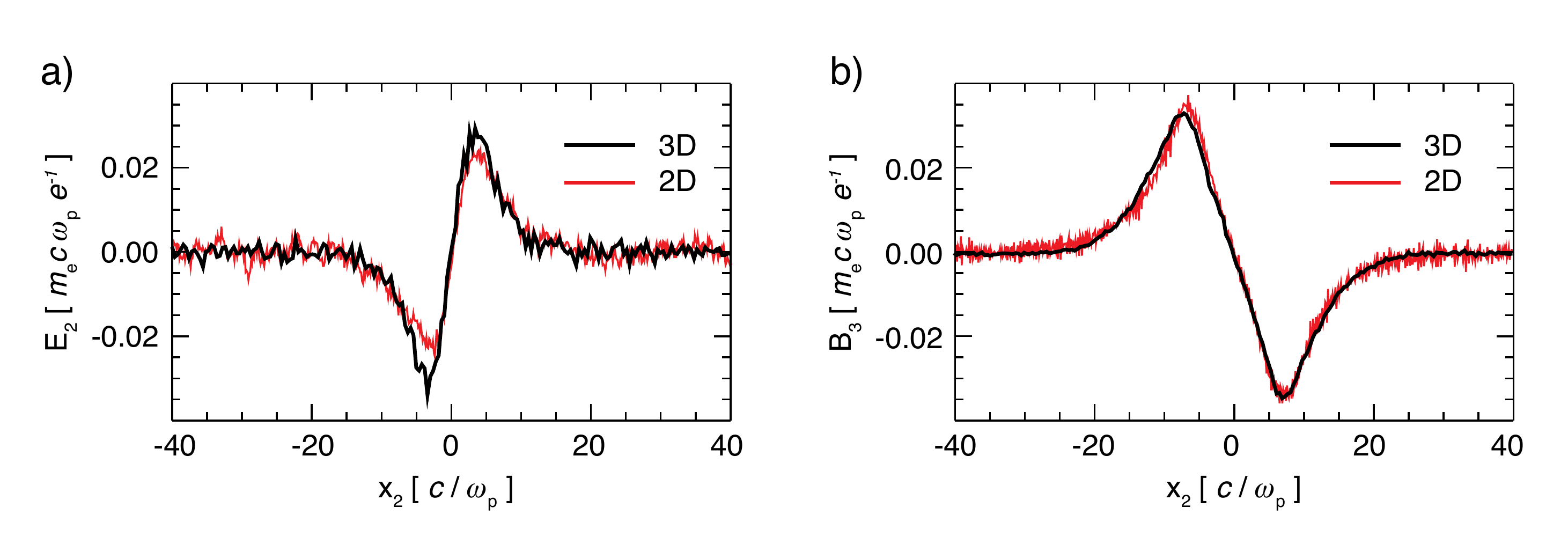}
	\caption{Comparison of the a) $E_2$ and b) $B_3$ components in $x_1-x_2$ plane for 2D and 3D simulations of linearly polarised laser channeling. Both panels represent lineouts for $x_1=345~c/\omega_p$ at $t=1632~\omega_p^{-1}$. } 
	\label{fields_2d3d}
\end{figure*}
 
Here we are interested in the electromagnetic field structure that is formed within a mm-scale plasma channel. 
This field structure can be used to aid particle acceleration or to guide an externally or self-injected electron beam.
To illustrate the channel field structure that could be expected in near-critical plasmas, we  perform a series of 2D and 3D PIC simulations of channel formation with OSIRIS \cite{Fonseca_scaling}. 
 The laser pulse length is 15 ps at FWHM, with the transverse laser waist of $W_0=$14.3 $\mu$m. 
 The temporal envelope consists in a 5 ps rise, 5 ps fall and a 10 ps flat-top section in between, where the laser amplitude is at its maximum. 
Peak laser intensity is $I=10^{18}~\mathrm{W/cm^2}$, corresponding to the normalized vector potential $a_0=0.8$. 
 The background plasma density is $n_e = 0.001~n_c$ which for a laser with $\lambda_0=1~\mu$m corresponds to $n_e=10^{18}~\mathrm{cm^{-3}}$. 
 The plasma is 1.91 mm long. 
 We simulate 31.9 ps of interaction with  $2.61\times10^{5}$ iterations. All simulations are normalized to $\omega_p=1.8\times10^{14}~\mathrm{s^{-1}}$. 
 With this normalization, distances expressed in $c/\omega_p$ can be radily converted to $\mu$m simply by multiplying with a numerical factor of 1.6.  
The simulation box in 2d is $2069~ \mu$m long and $159~\mu$m wide, resolved with $39000 \times 3000$ cells. 
The 3D simulations are performed with a box size of $2069~ \mu\mathrm{m} \times 127.4~ \mu\mathrm{m} \times 127.4~ \mu\mathrm{m}$, resolved with $39000 \times 200\times 200$ cells. 
We perform additional 2D simulations with $n_e=0.1~n_c$ and $n_e=0.01~n_c$, that are not possible in 3D because a finer transverse resolution would be required to do so. 
All 2D simulations are performed for a linearly polarized laser pulse, while in 3D we also consider circularly polarized laser pulses (considering the same $a_0=0.8$).

 \begin{figure*}
 	\centering
 	\includegraphics[width=1.0\textwidth]{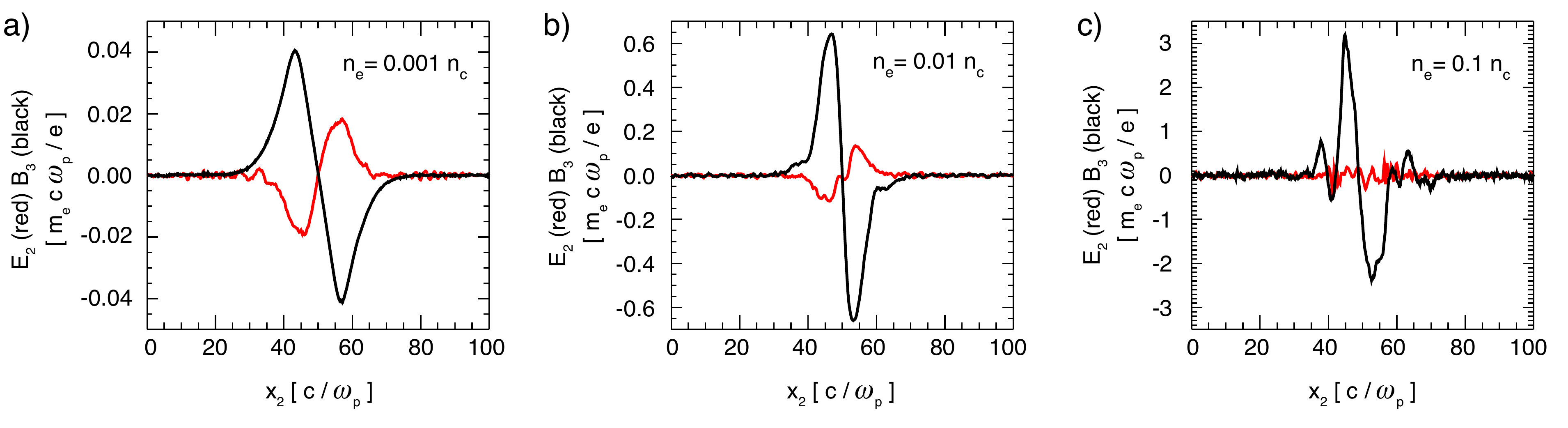}
 	\caption{Lineouts of space-charge electric field $E_2$ and out-of-plane magnetic field $B_3$ for plasma densities of a) $n_p=0.001~n_c$, b)  $n_p=0.01~n_c$ and c)  $n_p=0.1~n_c$. All lineouts are taken at $t=1150 ~\omega_p^{-1}$ (which corresponds to $\sim$ 6.4 ps) for $x_1=120 ~c/\omega_p$.} 
 	\label{EMfields_lineouts}
 \end{figure*}

  \begin{figure*}
 	\centering
 	\includegraphics[width=1.0\textwidth]{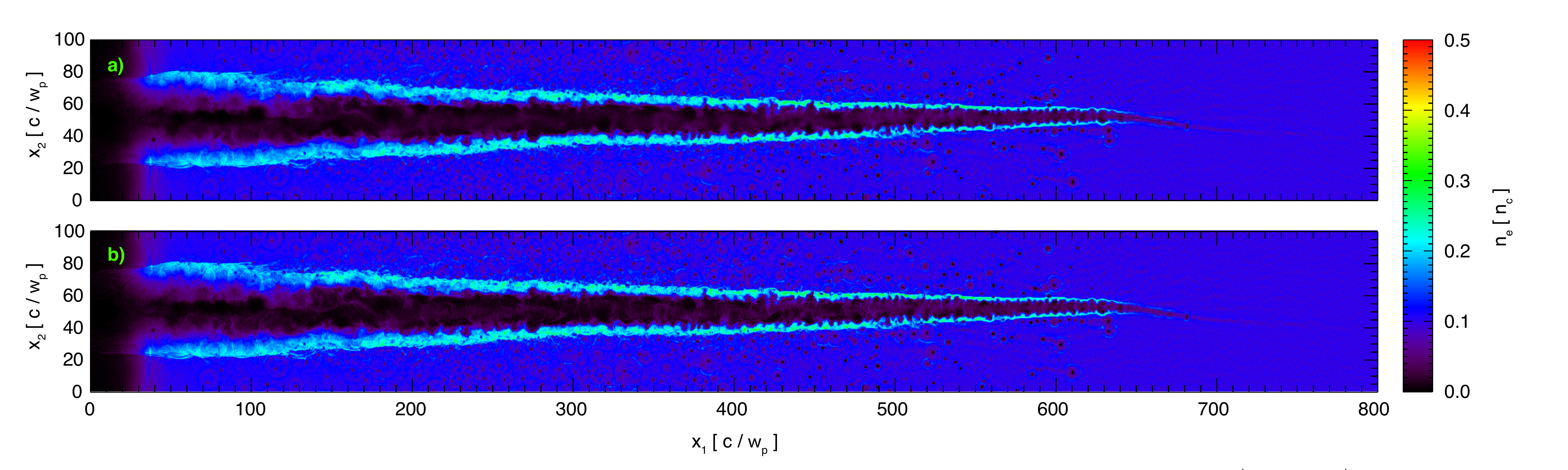}
 	\caption{Channel formation in a plasma with a background density of $n=0.1~n_c$. a) Electron plasma density, b) ion plasma density at $t=2300~\omega_p^{-1}$, which corresponds to $\sim$ 12.8 ps. } 
 	\label{long_channel_density}
 \end{figure*}

Figure \ref{3d_field_lines} shows the electromagnetic fields within the channel from the 3D simulations for circular and linear laser polarisations. 
The channel fields are superimposed with the fields of the laser, but are more than two orders of magnitude smaller. 
The fast-oscillating laser fields therefore interfere with measuring the slow-varying large-scale channel fields. 
However, we can still access the channel fields if we average the data over the fast oscillating component in the $x_1$ direction. 
It then becomes evident that the channel field structure is composed of a radial electric field and an azimuthal magnetic field, which are shown in Fig. \ref{3d_field_lines}. 
Let us now consider an electron beam traveling in the positive $x_1$ direction, and the effect the channel fields have on its propagation.
The radial electric field provides for the electrons a restoring force that acts towards the channel centre.
The direction of the azimuthal $B$ is also such that for an electron moving forward in $x_1$ in the vicinity of the channel axis,  the Lorentz force points towards the channel centre. 
Hence, both the electric and the magnetic fields of the channel provide forces that tend to guide the electron beam by acting against the escape of individual electrons from the central region.
This ensures a stable propagation of an electron beam or a current fillament within the channel. 

Full-scale 3D simulations of the channel formation process in 3D are possible only at low densities ($n_e\sim0.001~n_c$). 
By comparing 2D and 3D description at low densities, we evaluate how well laser channeling is described in 2D simulations.  
Figure  \ref{fields_2d3d} shows $E_2$ and $B_3$ lineouts parallel to the $x_2$ axis through the centre of the channel (here the $x_2$ coordinate represents the distance from the channel centre). 
One notable difference is that the electric field amplitude is slightly lower in 2D (as shown in \ref{fields_2d3d} a)). 
Apart from this, the field structure is identical in 2D and 3D, and the channel expands to the same width.  

For higher plasma densities, we deduce the order of magnitude of the fields generated, as well as the transverse size of the plasma channel from 2D simulations. 
The electric and magnetic field lineouts from for background plasma densities up to $n_e=0.1~n_c$ are given in Fig. \ref{EMfields_lineouts}. 
It is clear that denser plasmas generate proportionally higher space-charge fields in otherwise identical conditions. 
However, Fig. \ref{EMfields_lineouts} shows that this is true only for the magnetic field. 
The electric field does not scale linearly with the background plasma density. 
The reason is that the ions follow the electron transverse expansion on the channel formation timescale.
In fact, Fig. \ref{long_channel_density} shows that electron and ion density distributions are similar.
It is therefore not granted that by increasing the overall density one increases the electric field within the channel.  
Due to the existence of complex local structures, for the high-density background plasma with $n_e=0.1~n_c$, electric field fluctuations within the channel are on the same order as the expected sinusoidal space-charge electric field (as shown in Fig \ref{EMfields_lineouts} c).

Our simulations are in agreement with the charge displacement electric and the azimuthal magnetic fields within a laser-created channel that have already been observed experimentally using  proton probes \cite{Satya_channel, Ashley_Bfield}. 
Similarly, a strong azimuthal magnetic field is expected also during interaction with very intense lasers \cite{Arefiev_PRL2016_gamma_rays}. 
In the next section we  use the data from our simulations to define the channel properties for the case of $n_e=0.1~n_c$ and use it as an initial plasma configuration to propagate  extremely intense laser pulses. 

 \section{Extreme intensity laser channeling. The role of radiation reaction for particle trapping.} 
 
Now that we have determined what kind of channels  can be created with a 10 ps-class laser pulse, we investigate how this structure can serve to guide ultra-intense laser pulses  ($a_0\gtrsim100$). 
Due to the high intensity, it is expected that radiation reaction will significantly affect the plasma dynamics. 
We perform a series of 2D simulations at a range of laser intensities between $a_0=50$ and $a_0=600$.
Laser temporal profile has a 1.062 ps flat-top section with a 265 fs Gaussian rise and fall.
Background plasma density outside of the channel is $n_e=0.1~n_c$, the peak density of the channel walls is $n_e=0.2~n_c$ and the lowest density on channel axis is $n_e=0.02~n_c$. 
The channel width in the beginning of the simulation is 25.5 $\mu$m. 
The initialization of the pre-formed channel respects the values obtained in the previous section for $n_e=0.1~n_c$.
The simulation box is $1.27~ \mathrm{mm} \times 0.16~\mathrm{mm}$, resolved with $24 000 \times 3000$ cells. 
Total simulation time is 4.2 ps, with a $\Delta$t = 0.12 fs. 
Radiation reaction is described through classical Landau\&Lifshitz equation of motion \cite{Vranic_ClassicalRR}. 
To acertain this is  an adequate approach, we have verified that in our simulations the quantum  parameter $\chi_e<0.2$ even for most energetic particles in the simulations. 
The parameter $\chi_e$ is defined by $\chi_e=\sqrt{(p_\mu F^{\mu \nu})^2}/(E_c m_e c)$, where $E_c=m_e^2c^3/(\hbar e)$ is the Schwinger critical field. 
For particles counter-propagating with the laser, $\chi_e\simeq2a_0\gamma \times \hbar\omega_0/(m_ec^2)$, while for particles moving in the same direction as the laser $\chi_e \simeq a_0/(2\gamma) \times \hbar\omega_0/(m_ec^2)$.
Classical description is valid as long as $\chi_e\ll 1$.

\begin{figure}[t!]
	\centering
	\includegraphics[width=1.0\textwidth]{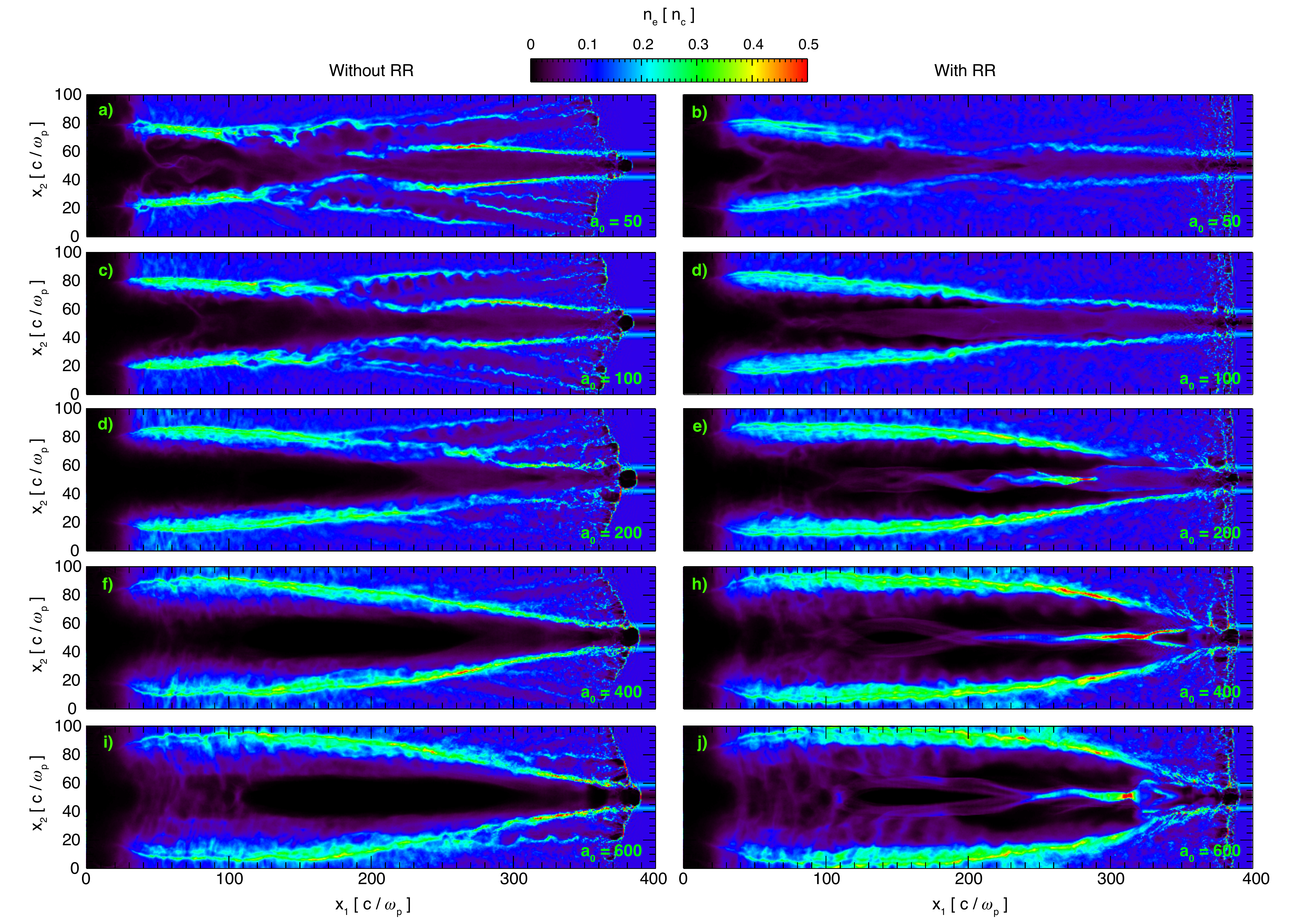}
	\caption{Channel density at $t=402.5~\omega_p^{-1}$ for extreme channeling simulations a), c), d), f), i) without and b), d), e), h), j) with radiation reaction. For $a_0>100$, apart from inhibited channel splitting with radiation reaction, there is a population of electrons within the channel that does not appear when radiation reaction is not accounted for. These electrons are in the region of the strongest electromagnetic field and can be accelerated.  } 
	\label{density_extreme}
\end{figure}

 Fig. \ref{density_extreme} shows the electron density for several simulations performed with and without radiation reaction (RR). 
 A striking difference in the collective plasma dynamics with and without RR is observed already for $a_0=50$. 
 Without RR, the channel bifurcates, which is not the case when radiation reaction is taken into account. 
 At higher laser intensities ($a_0>100$), there is a trapped electron beam in the vicinity of the channel axis in simulations with RR, that can be accelerated directly by the laser.
 When simulations are performed without radiation reaction, all the electrons are evacuated and there is no electron beam in the centre. 
 However, some direct laser acceleration occurs also without the radiation reaction, but away from the channel axis where the laser intensity is not at the maximum.  
 
\begin{figure}
	\centering
	\includegraphics[width=1.0\textwidth]{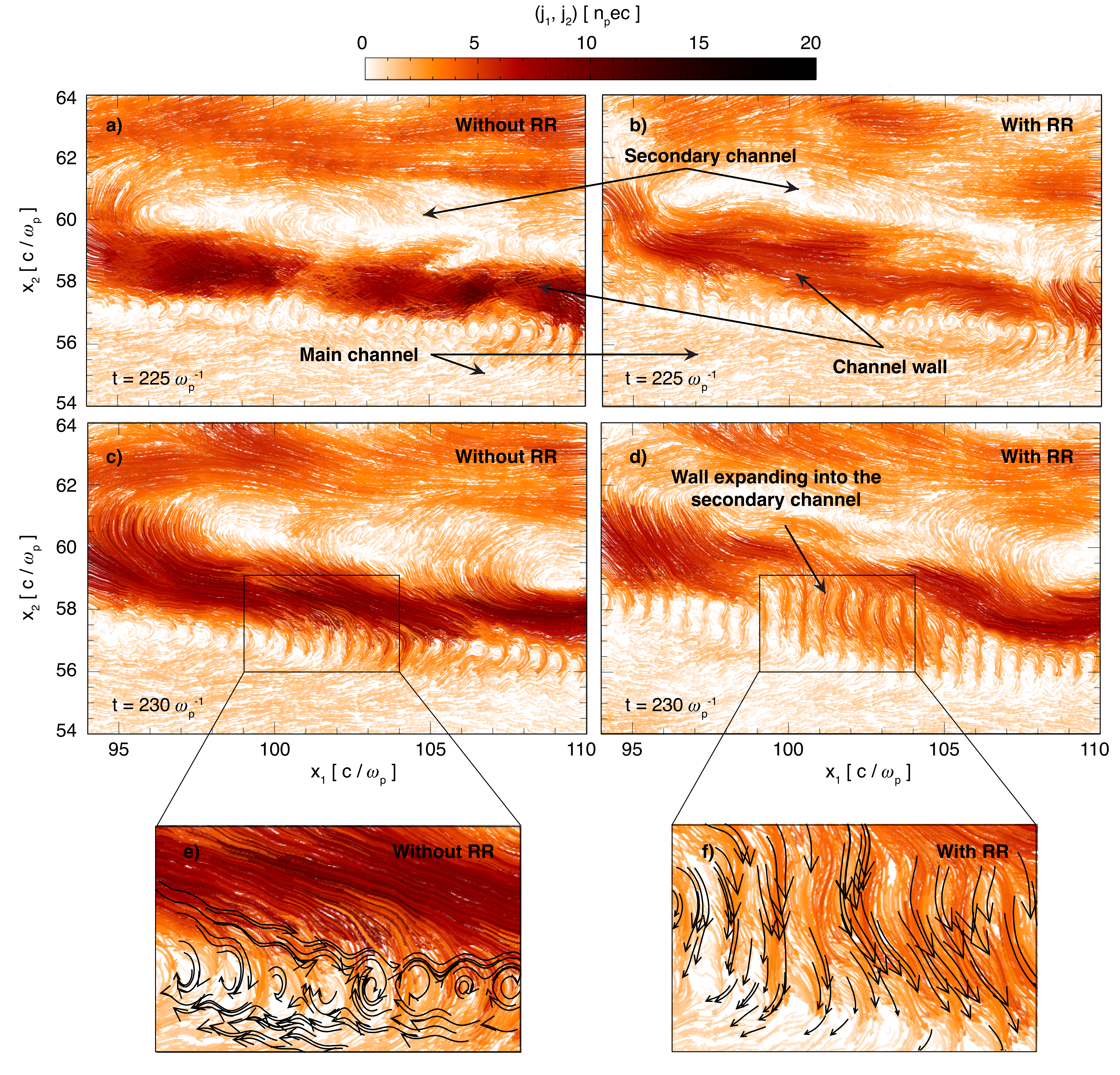}
	\caption{Vector plot of in-plane electron current a), c), e) without and b), d), f) with radiation reaction. The three frames show a region where the first channel splitting later occurs without RR, and does not occur with RR. Region around the channel wall e) without and f) with RR. Please note the direction of motion for  negatively charged electrons is opposite to the current vectors displayed. } 
	\label{current_vector}
\end{figure}

 We first investigate the channel bifurcation. 
 For a laser intensity coresponding to $a_0=50$, radiation reaction is not expected to strongly affect particles which are co-propagating with the laser. 
 This hints the radiation reaction is important near the channel wall, because that is the only region where there is a current flowing in the oposite direction. 
 To examine this in more detail, we focus on a region where the channel-splitting occurs in the simulation without radiation reaction. 
Figure \ref{current_vector} shows in-plane electron current precisely in this region right before the separation of a first side channel in the positive $x_2$ direction. 
Panels a) and b) in Fig. \ref{current_vector} show a similar current structure with and without RR at $t=225~\omega_p^{-1}$. 
Current loops are formed on the inside of the channel in the immediate vicinity of the channel wall. 
The existence of the loops near the channel wall is typical in near-critical plasmas \cite{Louise_channel}, and it preceeds channel splitting. 
The periodicity of these structures is on the order of the laser wavelength (the channel walls reach densities close to $n_c$, therefore the electron plasma frequency is also on the same order in this region). 
Qualitative difference between the two simulations appears already at $t=230~\omega_p^{-1}$. 
In the simulation without radiation reaction the current loops persist, while with radiation reaction they become elongated. 
The reason is that in the upper part of the current loop in Fig.  \ref{current_vector} e), the electrons move opposite the laser propagation direction. 
Due to the interaction with the laser, some electrons may be reflected forward. 
The reflection could happen with and without radiation reaction, this depends only on the local laser intensity and the counter-propagating particle energy.
However, the instantaneous energy of the electrons with RR is lower than without RR in otherwise identical conditions, because a fraction of electron energy is lost due to the radiation emission. 
This is what makes the reflection more probable with radiation reaction \cite{capturePiazza}. 
As their momentum already has an $x_2$ component, the reflected particles continue to propagate away from the channel axis as in Fig. \ref{current_vector} f).
The particles eventually close the current loop where the local laser intensity is lower.   
The elongated current loops contribute to forming a less sharp channel boundary which helps re-connecting the main and the side channel. 
The conclusions above apply also to lasers of higher intensities as the split channels consistently appear in the simulations without radiation reaction, but not in simulations with radiation reaction (Fig. \ref{density_extreme}). 

\begin{figure}
	\centering
	\includegraphics[width=0.8\textwidth]{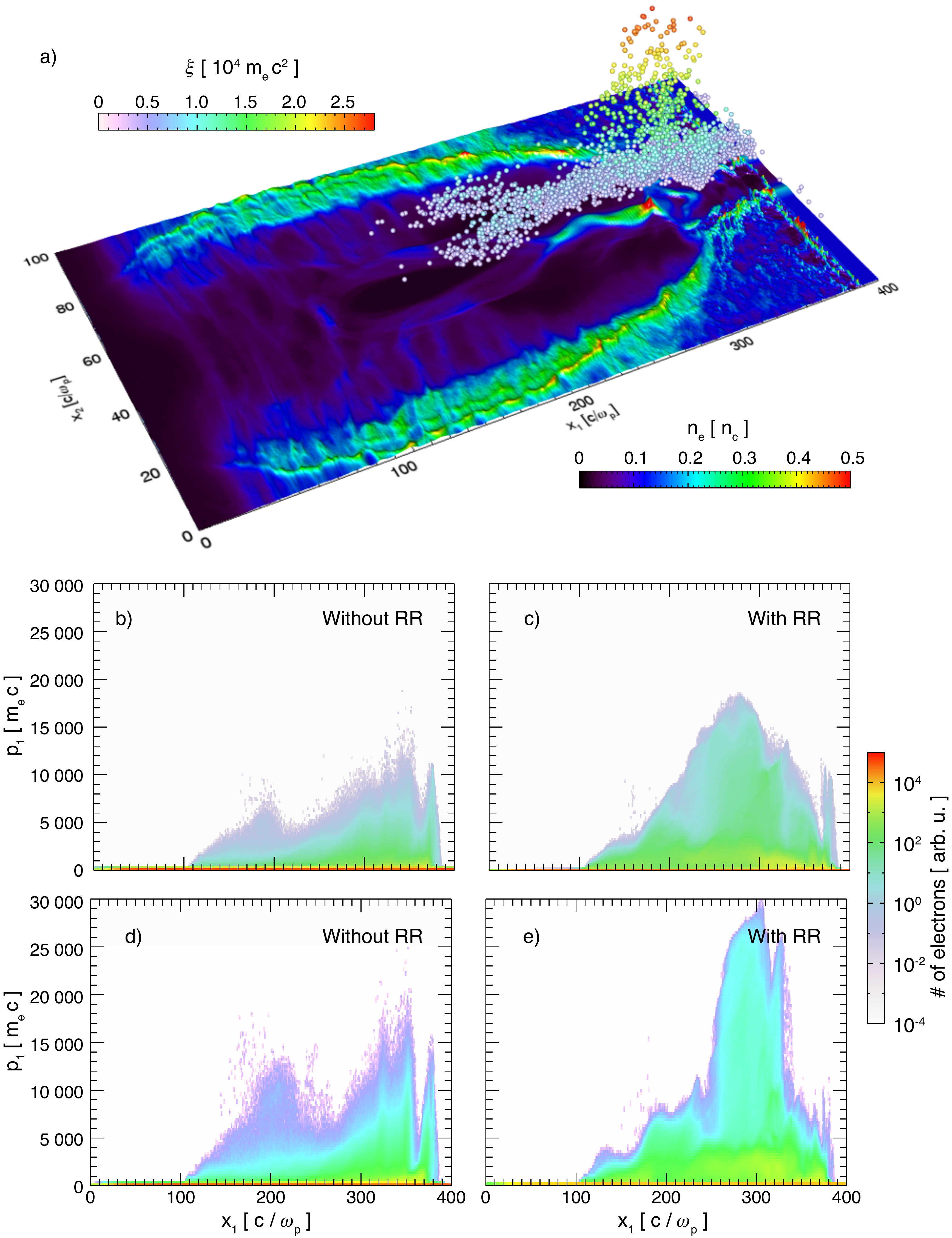}
	\caption{Electron beam energy at $t=402.5~\omega_p^{-1}$. a) Density of the channel for $a_0=600$ with randomly electrons with relativistic factor $\gamma>2000$. Spheres that represent individual electrons are coloured according to the energy. Their vertical distance from the $x_1-x_2$ plane also corresponds to  the energy. Most of these electrons are located within the channel in the central region that experiences the strongest laser field. b), c) Longitudinal phasespaces with and without radiation reaction respectively for $a_0=400$.  d), e) Longitudinal phasespaces with and without radiation reaction respectively for $a_0=600$. In general, the maximum electron energies obtained with RR are higher. } 
	\label{phasespaces_extreme}
\end{figure}

Apart from the inhibited channel splitting, another important difference is that in the simulations with RR for $a_0>100$ the channel has a dense population of electrons near the central axis. 
This threshold intensity for radiative trapping is of the same order, but slightly lower than in Ref. \cite{Pukhov_rrtrapp} that considered a shorter laser pulse. 
A radiation reaction strong enough to compensate for the laser expelling force was observed, which allowed the electrons to remain in the region close to the peak laser field. 
The electrons within the channel in these conditions can  be accelerated to energies  above 10 GeV, as illustrated in Fig.  \ref{phasespaces_extreme}.
Panel a) shows the channel density map together with randomly selected electrons with relativistic factor $\gamma>2000$. 
The laser intensity here is $a_0=600$, and all the electrons above 1 GeV reside in the vicinity of the channel axis.  
As the channel is not fully cavitated, the particles are accelerated both with and without radiation reaction. 
However, the most energetic particles are located at the channel axis, and these were injected there by radiative trapping. 
The longitudinal phase spaces in Fig. \ref{phasespaces_extreme} b)-e) show that  higher  longitudinal electron momenta are achieved with radiation reaction. 
The peak of high-energy electrons close to $x_1=300$ in panel e) corresponds to the very central region of the bunch close to the channel front. 
Therefore, assisted by the radiation reaction, intense lasers can produce energetic and collimated electron beams within hollow channels.  

\begin{figure}
	\centering
	\includegraphics[width=0.5\textwidth]{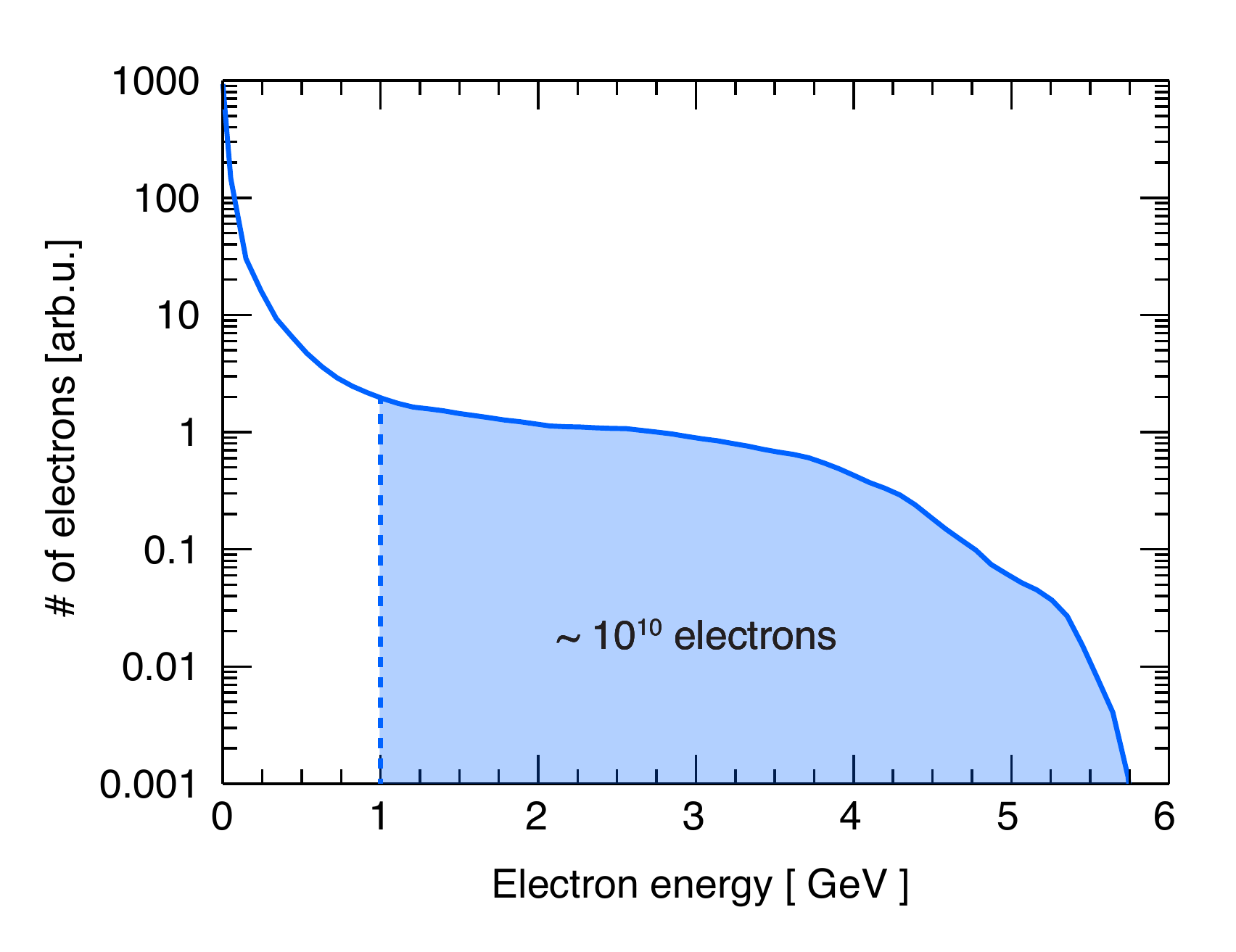}
	\caption{Electron  energy spectrum from a simulation with a  10 PW laser beam with a pulse duration of 150 fs (soon to be available at ELI \cite{facilities_ELI}). The highlighted section of the spectrum corresponds to $\sim$1.6 nC of charge. The spectrum is recorded after 1.8 mm of laser propagation. } 
	\label{electron_spect}
\end{figure}

Figure \ref{electron_spect} shows the electron spectrum of a beam accelerated in an identical plasma channel as above, but with a 150 fs long, 10 PW laser beam (to be available in ELI beamlines \cite{facilities_ELI}). 
The laser was initially focused to a $5~ \mu$m FWHM focal spot.
The electron energy cutoff is 6 GeV, with an equivalent of more than $10^{10}$ electrons accelerated above 1 GeV. 
The beam divergence is on the $50$ mrad level.
All these electrons are super-ponderomotive, as the laser energy is distributed within the channel and most of the propagation time the peak intensity corresponds to $a_0\sim50$.
During the propagation through 1.8 mm of plasma, the laser has lost less than 25 \% of its total energy and therefore the acceleration could potentially continue further within a longer plasma channel.
Using a channel with a smaller transverse size would be favourable to maintain a high laser intensity over a longer propagation distance. 
This would ultimately result in a higher beam energy cutoff and increase the amount of accelerated charge. 

\begin{figure*}
		\centering
		\includegraphics[width=1.0\textwidth]{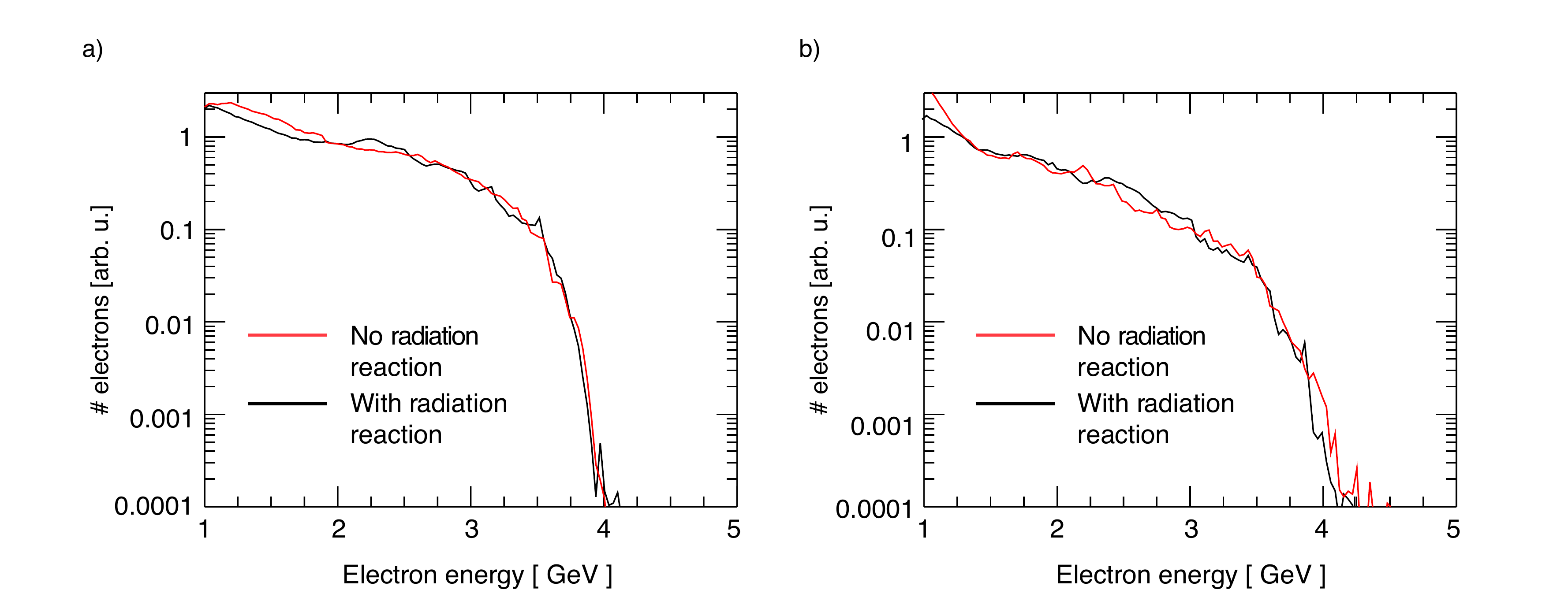}
		\caption{Electron energy spectrum with and without radiation reaction. a) For a constant channel width of 25.5 $\mu$m; b) the channel width varies in $x_1$ direction as in Fig. \ref{long_channel_density}.} 
		\label{spec_compare}
\end{figure*}

One question that might arise is how a spatio-temporal variation of the channel width could affect the results above. 
As the channel expands at the local sound speed, the width variations are small within the $\sim150$ fs timescale and we assume the channel does not evolve during the interaction. 
However, spatial variation (in $x_1$ direction) depends on the first pulse parameters and the  delay before the main pulse arrives. 
We can map the varying channel width from Fig. \ref{long_channel_density}, where the channel is $\sim 32 ~ \mu$m wide at the vacuum-plasma interface and $\sim9 ~\mu$m wide 1 mm into the plasma. 
Figure \ref{spec_compare} shows two electron spectra obtained with a constant channel width above, and using a channel as in Fig. \ref{long_channel_density}. 
The energy cutoff is similar, but within the spatialy varying channel 30 \% fewer electrons were accelerated above 1 GeV in the first mm of propagation. 
For completeness, we have displayed the spectra with and without radiation reaction, even though at $a_0\sim50$ the differences in the accelerated beam are minor.

We note that the configuration presented here is the simplest example where radiative trapping is of significance at extreme laser intensities. 
The effect can also be found in standing waves formed by interaction of two or more lasers  \cite{Gonoskov_radiative_trapping, Kirk_radiative_trapping}, which is beyond the scope of this paper.
We stress, however, that the configuration explored here can be tested in the next generation of laser facilities such as ELI \cite{facilities_ELI}.


\section{Conclusions} \label{concl_sect}
	
We have studied channel formation and intense laser propagation through pre-formed channels. 
We have shown that prepulses can be exploited to generate parabolic channels that would serve as a light pipe to guide the intense lasers through large-scale underdense plasmas. 
Electrons with energies above 10 GeV can be obtained within a 1-mm long plasma channel using the next generation of laser facilities \cite{facilities_ELI}. 
By focusing a 10 PW laser pulse with a duration of 150 fs to a $5~ \mu\mathrm{m}$ focal spot, 6 GeV energy gain can be obtained in 1.8 mm of propagation. 
The channel width used here is $25.5~\mu\mathrm{m}$. 
This was not in any way optimized to maintain the maximum laser intensity or generate maximum electron energy.  
Further studies are required to show the optimal conditions for the proposed acceleration setup.
This would likely increase the electron acceleration efficiency and the maximum energy the electrons obtain in the light pipe.

	
	
	\section*{Acknowledgements}
	This work is supported by the European Research Council (InPairs ERC-2015-AdG Grant 695088), and FCT (Portugal) SFRH/BPD/119642/2016. Simulations were performed at Supermuc (Germany) and Fermi (Italy) through PRACE allocation and at the IST cluster (Lisbon, Portugal). The authors thank Dr T. Grismayer for fruitful discussions. 
	
	\section*{References}
	\bibliographystyle{unsrt}

\end{document}